\documentclass[12pt]{iopart}

\expandafter\let\csname equation*\endcsname\relax
\expandafter\let\csname endequation*\endcsname\relax
\usepackage{amsmath}
\usepackage{bm}
\usepackage{graphicx}
\usepackage{amsfonts}
\usepackage{amssymb}
\usepackage{units}
\usepackage{textcomp}
\usepackage[ansinew]{inputenc}
\usepackage{gensymb}
\usepackage{color}

\newcommand{\C}{$\text{C}_{60}$}
\newcommand{\hex}{$h$-$\text{C}_{60}$}
\newcommand{\pen}{$p$-$\text{C}_{60}$}

\begin{document}

\title{Force and conductance during contact formation to a \C\ molecule}

\author{Nadine Hauptmann$^1$, Fabian Mohn$^2$, Leo Gross$^2$, Gerhard Meyer$^2$, Thomas Frederiksen$^3$ and Richard Berndt$^1$}
\address{$^1$Institut f\"{u}r Experimentelle und Angewandte Physik, Christian-Albrechts-Universit\"{a}t zu Kiel, D-24098 Kiel, Germany\\}
\address{$^2$IBM Research -- Zurich, CH-8803 R\"{u}schlikon, Switzerland\\}
\address{$^3$Donostia International Physics Center (DIPC), E-20018 Donostia-San Sebasti$\acute{a}$n, Spain}
\ead{hauptmann@physik.uni-kiel.de}

\begin{abstract}
Force and conductance were simultaneously measured during the formation of Cu--\C\ and \C--\C\ contacts using a combined cryogenic scanning tunneling and atomic force microscope.
The contact geometry was controlled with submolecular resolution.
The maximal attractive forces measured for the two types of junctions were found to differ significantly.
We show that the previously reported values of the contact conductance correspond to the junction being under maximal tensile stress.
\end{abstract}


\maketitle

\section{Introduction}
When a molecule is contacted by electrodes to measure the conductance of the molecular junction, new bonds are formed and significant forces may arise.
These forces affect the atomic-scale junction geometry, which is crucial for its transport properties ~\cite{Cross1998,AGR,Limot2005,Chen2007,Neel2007}.
Current and force can be measured simultaneously using a combination of scanning tunneling microscopy (STM) and atomic force microscopy (AFM).
Such measurements were carried out for metallic contacts~\cite{AGR,DRG,SUN,Sawada2009,Ternes2011}.
Related data was reported for contacts to single molecules in a liquid environment~\cite{Xu2003,Ebeling2009} and for molecules on a metal surface~\cite{Lantz1999}.
However, the exact contact geometry was not accessible.
3,4,9,10-perylene-tetracarboxylicacid-dianhydride was probed in ultrahigh vacuum using AFM to controllably lift the molecule~\cite{Fournier2011}.
A bimodal distribution of conductances was observed and suggested to reflect two distinct bonding geometries.
As to controlled molecule--molecule contacts, experimental results are few.
The conductance of \C--\C\ contacts was measured by attaching a \C\ molecule to an STM tip and approaching a second molecule in a  monolayer on Cu(111)~\cite{Schull2009}.
The force between a metal tip and \C\ molecules in double layers on Cu(111) was addressed with AFM \cite{Pawlak, Pawlak2}.
While close distances well into the repulsive range were explored, the corresponding conductances \cite{instvsave} were significantly lower than in the STM work of Ref.~\cite{Schull2009}.
A possible origin for this difference may be different geometries of the contact between the tip and the molecule in these experiments.
Atomically sharp electrodes were shown to act as bottlenecks for charge injection into \C~\cite{Schull2011nn,Schull2011}.
While tips had been intentionally flattened to firmly attach a molecule in Ref.~\cite{Schull2009}, the tip used in Ref.~\cite{Pawlak} presumably was atomically sharp.
Another possible reason for reduced conductance is foreign material at the tip apex.
Here, we present low-temperature force and conductance data for the controlled formation of Cu--\C\ and \C--\C\ contacts.
The orientations of the molecules at the tip and the surface were determined from STM imaging.
The elasticity of both contacts is analyzed and compared with density functional theory (DFT) calculations.

\section{Experiment}
Our experiments were performed with a homebuilt STM/AFM in ultrahigh vacuum at a temperature of $5\,\text{K}$.
Clean Cu(111) surfaces  were prepared by repeated sputtering and annealing cycles.
Submonolayer amounts of \C\ were then deposited onto the sample by sublimation at room temperature. Subsequent annealing to $\approx 500\,\text{K}$ led to a well-ordered $4\times 4$ structure of \C~\cite{Hashizume1993,Pai2004,Wang2004,Pai2010}.
After additional sublimation of small amounts of \C\ onto the cooled sample, isolated \C\ molecules were found on both the \C\ islands and the bare Cu substrate~\cite{Larsson2008}.
A PtIr tip was attached to the free prong of a quartz tuning fork oscillating with an amplitude of $(3\pm0.2)\,\text{\AA}$ at its resonance frequency of $\sim 28\,\text{kHz}$.
The tip was covered with Cu by repeated indenting into the substrate until submolecular resolution was achieved.
The vertical force $F$ acting on the tip at the point of closest tip approach was calculated from the measured frequency shift $\Delta f(\Delta z)$ (shown in S1) as a function of the vertical piezo displacement $\Delta z$ using the formalism of Sader and Jarvis~\cite{Sader2004}.
Due to the limited bandwidth of the transimpedance amplifier, the current recorded with the oscillating tip is averaged over the entire range of oscillation.
The non-averaged conductance $G(\Delta z)$ was calculated using the method of Sader \emph{et al.}~\cite{Sader2010}, which recovers the instantaneous current at the point of closest tip approach.
The bias voltage $V$ was applied to the sample.
Further experimental details can be found in the supplementary data.

We note that the intrinsic energy dissipation of the tuning fork did not change significantly during the contact formation.
In addition, STM images taken before and after the contact measurement showed no changes.
These facts suggest that no inelastic deformations of tip or molecule occurred.

\section{Cu--\C\ contacts}

\begin{figure}
\centering
\includegraphics[width=120mm,clip=]{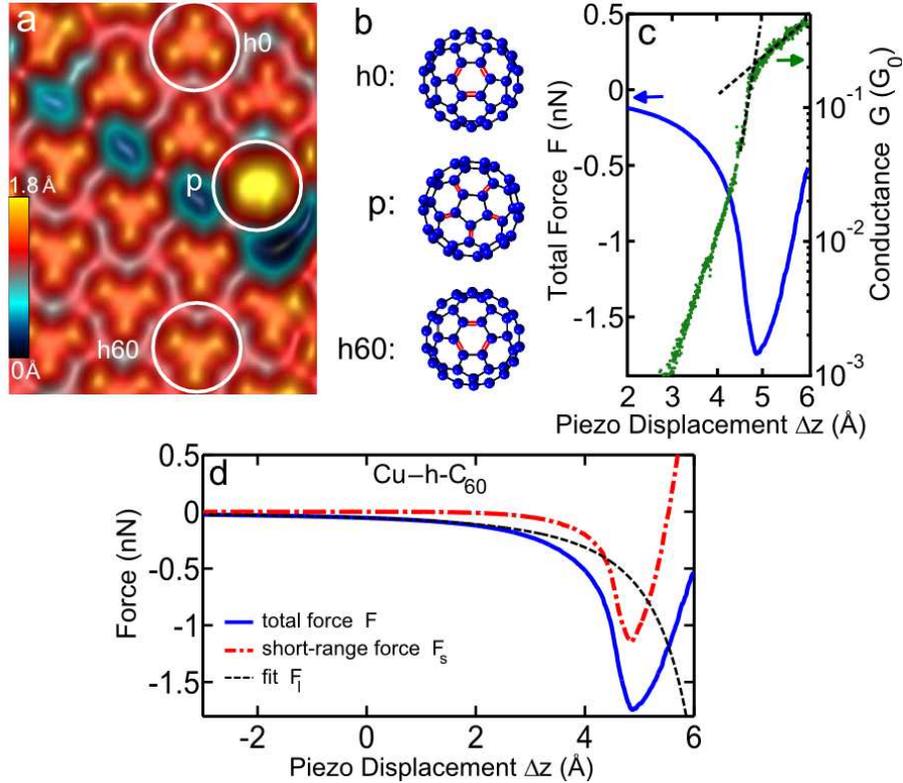}
\caption{(a) Pseudo-three-dimensional illuminated constant-current STM image ($1.7\,\text{V}$, $0.55\,\text{nA}$, $3.8\times 5\,\text{nm}^2$) of a \C\ island showing three different molecule orientations (h0, p, and h60).
(b) Sketches of three orientations of \C\ on Cu(111) as viewed from the tip.
Red double-bars indicate the bonds separating two hexagons (6:6 bonds).
(c) Total force $F$ (solid line) and instantaneous conductance $G$ (dots) calculated from the simultaneously measured frequency shift and averaged conductance (Fig.~S1) using Refs.~\cite{Sader2004,Sader2010}. A voltage of $V =0.1\,\text{V}$ was applied to the sample. The oscillation amplitude of the tuning fork was $A=(3\pm0.2)\,\text{\AA}$. Dashed lines define the point of contact.
(d) Total force $F$ (solid line) over a wider range of piezo displacements than in (c), fitted long-range force $F_l$ (dashed line, fit range: $\Delta z\leq 0\,\text{\AA}$, see text for details) and short-range force $F_s=F -F_l$ (dashed-dotted line).
$\Delta z = 0\,\text{\AA}$ corresponds to the position defined by the STM set-point $I=0.55\,\text{nA}$ at $V=1.5\,\text{V}$.}
\label{fig:CuC60}
\end{figure}

Figure~\ref{fig:CuC60}(a) displays a typical constant-current STM image of a \C\ island used for contact measurements with Cu-covered tips.
The island comprises two domains which differ by an azimuthal rotation of the \hex\ molecules by 60\degree\ (h0 and h60).
\C\ molecules adsorbed on Cu(111) with either a hexagon (\hex) or a pentagon (\pen) facing up give rise to distinctly different patterns in the STM image \cite{Silien2004}.
Fig.~\ref{fig:CuC60}(b) shows the different orientations of the molecules as viewed from the tip position.
Double bonds separating two hexagons (6:6 bonds) are marked by red bars.
Contact data recorded with a Cu tip above the center of a \hex\ molecule are shown in Fig.~\ref{fig:CuC60}(c).
Both the total interaction force $F$ (solid line) evaluated at the point of closest approach to the sample and the instantaneous conductance $G$ (dots) are displayed versus the piezo displacement $\Delta z$.
We first focus on the force, which is shown over a wider range of displacements in Fig.~\ref{fig:CuC60}(d).
To minimize the electrostatic force, which results from the contact potential difference, a bias voltage of $V =0.1\,\text{V}$ was applied during the contact measurement~\cite{Nonnenmacher1991,Kitamura1999,Okamoto2002}.
At large tip--sample distances, $F$ reflects the long-range van-der-Waals force between the tip and the sample.
It can be approximated by a power law $F_l(\Delta z) = a (\Delta z_0 - \Delta z)^b$~\cite{Israelachvili1992} with typical fit parameters $a=-5.5\,\text{nN}/\text{\AA}^{b}$, $b=-2.3$, and $\Delta z_0=7.5\,\text{\AA}$ (fit range: $\Delta z \leq 0\,\text{\AA}$). The fit is shown in Fig.~\ref{fig:CuC60}(d) as a dashed line.
The exponent close to 2 indicates an effective sphere--plane geometry of the junction.

The short-range force $F_s(\Delta z)$ [dashed-dotted line in Fig.~\ref{fig:CuC60}(d)], which only acts on the atoms in the immediate vicinity of the molecular junction, is estimated as $F_s = F -F_l$.
It is attractive for large tip heights, reaches a minimum at $\Delta z \approx 4.8$\AA, and finally becomes repulsive for $\Delta z > 5.5\,\text{\AA}$.
The fitted van-der-Waals force at contact ($\approx 0.5$ nN) is consistent with estimates for a sharp tip \cite{hobbs2006}.
We note that the total force $F$ and the short-range force $F_s$ exhibit maximal attraction at nearly the same position $\Delta z.$
In other words, long-range forces do not significantly affect the point of maximal attraction.
However, they do affect the value of the maximal attraction.
Using different tips we found it to scatter between $1.5\,\text{nN}$ and $2.2\,\text{nN}$.
We ascribe the origin of these significant short-range forces to the chemical bond formation between tip and molecule. Interestingly, calculated interaction forces for a Si tip and \C\ on Si(100) are in a very similar range ($1.4 \dots 2.0\,\text{nN}$) \cite{hobbs2006}.
The $F(\Delta z)$ data of \hex\ and \pen\ were very similar except for a shift of $\approx 0.4\,\text{\AA}$ along the abscissa due to the different apparent heights of the molecules.
This apparent insensitivity to the detailed bonding geometry may be attributed to the high reactivity of 6:6 bonds.
It causes the Cu atom at the tip apex to laterally relax \cite{Schull2011}.
As a result, a 6:6 bond is most likely contacted independent of the orientation of the molecule.
The conductance $G(\Delta z)$ in Fig.~\ref{fig:CuC60}(c) shows a typical transition from tunneling at small $\Delta z$ to electrical contact.
To define the point of contact, the intersection of linear fits in the transition and contact regime is used [Fig.~1(c), dashed lines]~\cite{Neel2007}.
The resulting contact conductance $G_c \approx 0.2\,G_0$  ($G_0=2 e^2 / h$ is the conductance quantum) is in agreement with previous experimental results~\cite{Neel2007,Schull2009,Schull2011nn}.
Comparing the conductance data with the simultaneously measured force [Fig.~1(c), solid line] we find that the point of contact corresponds to maximal attractive force.
The same observation is made for \C--\C\ contacts (see below).
A similar result has been reported from metal--metal point contacts, where a maximal attractive force was measured at $G\sim G_0$~\cite{Ternes2011}.
Modeling of metallic contacts also suggests that the deformation of the junction is maximal at the point of contact formation \cite{Olesen1996,Trouwborst2008}. Recently, two-level fluctuations of the conductance on a $\mu$s time scale have been reported from \C\ on Cu(100) at the transition from tunneling to contact for a metal-\C\ contact\,\cite{Neel2011}.
In the present case of a $4\times 4$ structure of \C\ on Cu(111), and at the rather low bias voltages used ($0.1\,\text{V} \dots 0.3\,\text{V}$), such fluctuations were not observed\,\cite{Natalia}.

\section{\C\--\C\ contacts}
By approaching the tip sufficiently close, a single \C\ molecule was attached to the tip apex.
The orientation of such \C\ tips was determined by 'reverse imaging' on small Cu clusters which had been deposited before from the tip onto the bare Cu(111) surface~\cite{Schull2009,Schull2011nn}.
Constant-current images of such a Cu cluster recorded at $V=-2\,\text{V}$ reveal the second lowest unoccupied orbital (LUMO+1) of the molecule~\cite{Schull2009}.
Compared to normal STM images of the LUMO+1 [Fig.~\ref{fig:CuC60}(a)], they show a mirror image of the molecule~\cite{Supplement}.

\begin{figure}
\centering
\includegraphics[width=80mm,clip=]{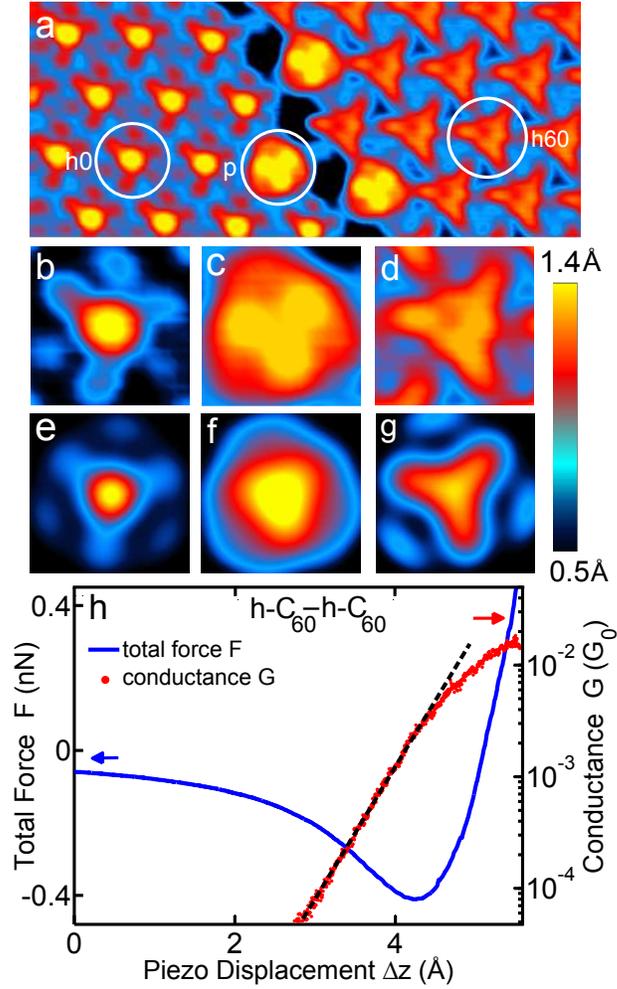}
\caption{(a) Constant-current STM image ($1.6\,\text{V}$, $0.55\,\text{nA}$, $8\times 4\,\text{nm}^2$) of a \C\ island measured with a \hex\ tip.
(b--d) Constant-current closeup images of molecules with orientations h0, p, and h60, respectively.
(e--g) Simulated images showing the convolution of the modeled local density of states (LDOS) of the HOMO of the \C\ at the tip and the LUMO of the \C\ at the surface for h0, p, and h60 orientations, respectively~\cite{Supplement}.
(h) Instantaneous conductance $G$ (dots) and total force $F$ (solid line) for the \hex--\hex\ contact calculated from the simultaneously measured frequency shift and averaged conductance (Fig.~S1) using Refs.~\cite{Sader2004,Sader2010}. An oscillation amplitude $A=(2.5\pm0.2)\,\text{\AA}$ was used. A voltage of $V =0.3\,\text{V}$ was applied the the sample. The dashed line indicates an exponential fit to the conductance for small $\Delta z$.}
\label{fig:C60C60}
\end{figure}

The relative orientations of the tip and sample \C\ molecules strongly affect the conductance of the junction in the tunneling range.
Figure~\ref{fig:C60C60}(a) shows a \C\ island imaged using a \C-functionalized tip with a hexagon facing towards the surface (\hex\ tip).
Similar to Fig.\ \ref{fig:CuC60}(a), the island comprises two rotational domains of \hex\ (h0 and h60), as well as a few \pen\ molecules.
Owing to the different orientations, distinctly different patterns are observed with the \hex\ tip for h0, p, and h60 molecules [Figs.~\ref{fig:C60C60}(b--d)].
For example, the center of \hex\ appears either as a maximum (h0) or as a minimum (h60) in the STM image.
On \pen a threefold symmetry of the \hex\ tip is clearly discernible, which reflects the 5:6 bonds of the molecule at the tip.
The symmetries of these patterns can be understood from a convolution of the local densities of electronic states (LDOS) of the tip and the sample.
At $V=1.6\,\text{V}$ electrons essentially tunnel from the highest occupied molecular orbital (HOMO) of the \hex\ tip to the lowest unoccupied molecular orbital (LUMO) of the molecule at the surface.
A two-dimensional convolution of these orbitals according to the orientations given in Fig.\ S2 is shown in Figs.~\ref{fig:C60C60}(e--g)~\cite{Supplement}.
It reproduces the experimental data with the best agreement obtained for the h0 pattern.

Figure~\ref{fig:C60C60}(h) displays the force (solid line) and the conductance (dots) measured on a \hex\ molecule with a \hex\ tip at an applied voltage of $V =0.3\,\text{V}$.
Compared to the data from a Cu tip [Fig.~\ref{fig:CuC60}(c)], the maximal attractive force is smaller by a factor of 4\@.
In experiments with different \C\ tips, this maximal attractive force varied from $0.3\,\text{nN}$ to $0.4\,\text{nN}$.
In part, this scatter may be attributed to the uncertainty of the lateral tip position, which we estimate to be $\approx 10\%$ of a \C\ diameter.
The conductance measured for a \C--\C\ contact [Fig.~\ref{fig:C60C60}(h), dots] starts to deviate from a purely exponential behavior (dashed line) at a piezo displacement close to the position of the maximum of the attractive force.
The conductance at this point ($\approx 3 \times 10^{-3}\,G_0$) is approximately two orders of magnitude smaller than with a Cu tip~\cite{Schull2009}.
When we approached the tip further towards the surface until the total force exceeded $0.5\,\text{nN}$, a rotation of the \C\ molecule at the tip occurred.

\section{Comparison of elasticities of Cu--\C\ and \C\--\C\ contacts}
The forces at the junction cause atomic relaxations which affect the conductance.
Below, a simple model is used to estimate the deformation of the junction from the measured conductance and force data.
First, the conductance of a rigid junction is calculated as a function of the piezo displacement, $G_\text{rigid}^\text{theo} (\Delta z)$ \cite{Schull2009,Supplement}, which is shown in Fig.~\ref{fig:Theo}.
\begin{figure}
\centering
\includegraphics[width=90mm,clip=]{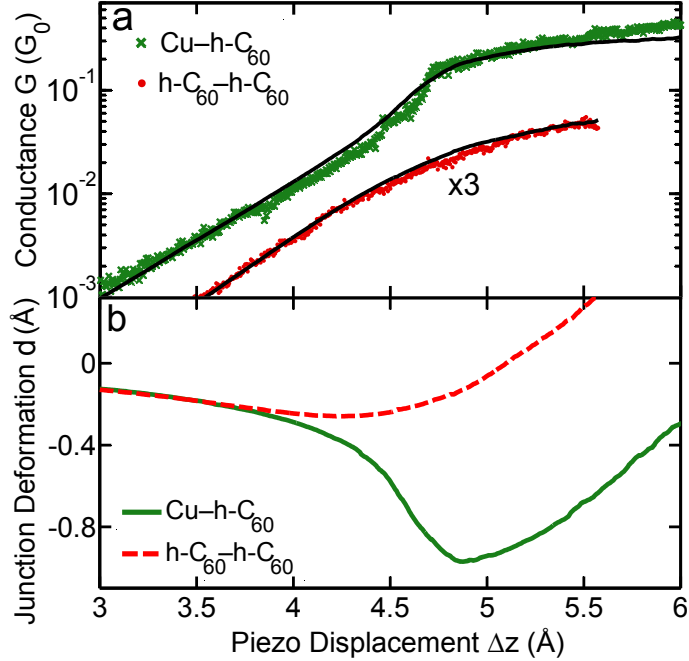}
\caption{(a) Conductance $G$ versus $\Delta z$ during the formation of Cu--\hex\ (crosses), and \hex--\hex\ (dots) contacts.
The solid lines are fits of $G_\text{deform}^\text{theo}$ to the measured conductance $G$ using Eq.\,(\ref{seq}) with $k=19.4 \,\text{Nm}^{-1}$ for Cu--\hex\ and $k=18.9\,\text{Nm}^{-1}$ for \hex--\hex. More details can be found in appendix B.
(b) The junction deformation $d$ extracted from the fits to $G(\Delta z)$ for Cu-\hex\ (solid line) and \hex--\hex (dashed line).}
\label{fig:Elast}
\end{figure}
The junction deformation $d$ is approximated by a linear relation $d = F(\Delta z)/k$, using the experimentally determined force $F(\Delta z)$.
We then obtain the theoretical conductance of the deformed junction, $G_\text{deform}^\text{theo} (\Delta z)$, which depends on the stiffness of the junction $k$:
\begin{equation}
G_\text{deform}^\text{theo} (\Delta z) =G_\text{rigid}^\text{theo} \Big(\Delta z-\frac{F(\Delta z)}{k}\Big)\,\,.\label{seq}\end{equation}
From a fit of $G_\text{deform}^\text{theo} (\Delta z)$ to the measured $G(\Delta z)$ data [Figs.\,\ref{fig:CuC60}(c) and \ref{fig:C60C60}(h)] the stiffness $k$ is obtained.
Figure~\ref{fig:Elast}(a) shows fits (solid lines) for Cu--\hex\ (crosses) and \hex--\hex\ (dots) contacts.
Cu--\pen\ data (not shown) yield similar results.
From measurements with different tips we determined elasticities $k\approx (16-37)\,\text{Nm}^{-1}$ for Cu--\C\ and $k\approx (13-24)\,\text{Nm}^{-1}$ for \C--\C.
The extracted deformation $d$ shown in Fig.\,\ref{fig:Elast}(b) corresponds to a reduction of the tip--molecule distance of $\approx 0.9\,\text{\AA}$ for the Cu--\C\ contact (solid line).
For the \C--\C\ contact (dashed line), $d$ is smaller ($\approx 0.26\,\text{\AA}$) and the transition from tensile to compressive deformation occurs within the $\Delta z$ range that was accessible in our experiment.
While the deformations are smaller than the values reported for metal contacts~\cite{Ternes2011,Olesen1996,Hofer2001}, they still significantly affect the conductance.

The values for $k$ may be interpreted in terms of the elasticities of the  components of the junctions.
DFT calculations taking into consideration several atomic configurations were used to estimate elasticities of tip and sample (see appendix A).
As summarized in Tab.~\ref{tab:springs}, we find that a metallic Cu tip can be characterized by a stiffness in the
range $k_\textrm{Cu}^\textrm{tip} \approx (45-55)\,\text{Nm}^{-1}$ depending on its atomistic structure and on details of the calculational scheme.
For the sample we find $k_\textrm{C60}^\textrm{sample} \approx (112-129)\,\text{Nm}^{-1}$ for \hex\ or \pen\ on reconstructed Cu(111)~\cite{Pai2010}.
Combining tip and sample elasticities in series, we thus estimate $k_\textrm{eff} \approx (32-39)\,\text{Nm}^{-1}$ for a Cu--\C\ contact.
Similarly, for a \hex-tip we find $k_\textrm{C{60}}^\textrm{tip} \approx (43-81)\,\text{Nm}^{-1}$, which leads to $k_\textrm{eff} \approx (31-50)\,\text{Nm}^{-1}$ for a \hex--\hex\ contact.
For both contacts, the experimentally determined elasticity is softer than the calculated one by a factor of 2\@.
We attribute this difference to two main factors:
First, the elasticity calculations can be considered as upper bounds as only a finite number of degrees of freedom are taken into account (see appendix A).
Second, the elasticity estimates above do not take into account the softening of the springs close to contact due to the formation of chemical bonds between tip and sample.

It is instructive to compare the obtained junction stiffness values with that of an isolated \C\ molecule.
From our DFT calculations we find that squeezing a \C\ molecule between two opposite hexagons corresponds to a spring constant of 222 N/m (appendix A), i.e., the elasticity of a \C\ is significantly stiffer than the molecular junctions considered in this study.
The junction deformation therefore mainly involves the metal-molecule bonds and the STM tip. 

\section{Conclusions}
In summary, simultaneous force and conductance measurements for Cu--\C\ and \C--\C\ contacts have been performed.
Angstrom-scale deformations of the contacts and effective stiffness extracted from the experimental data agree with elasticities determined with DFT calculations.
We find that the maximal attractive force measured at a \C--\C\ contact is 4 times smaller than in a Cu--\C\ junction.
Moreover, the force data reveal that previously reported contact conductances correspond to geometries in which the junctions are under maximal tensile stress.

\section*{Acknowledgments}

We thank C. Gonz\'alez and N. L. Schneider for discussions. Financial support by the Deutsche Forschungsgemeinschaft (SFB 677), the Innovationsfonds Schleswig-Holstein, and the EU projects HERODOT and ARTIST is acknowledged.

\section*{Appendix}
\appendix
\section{DFT calculations}

\begin{figure}[ht]
\centering
\includegraphics[width=\textwidth,clip=]{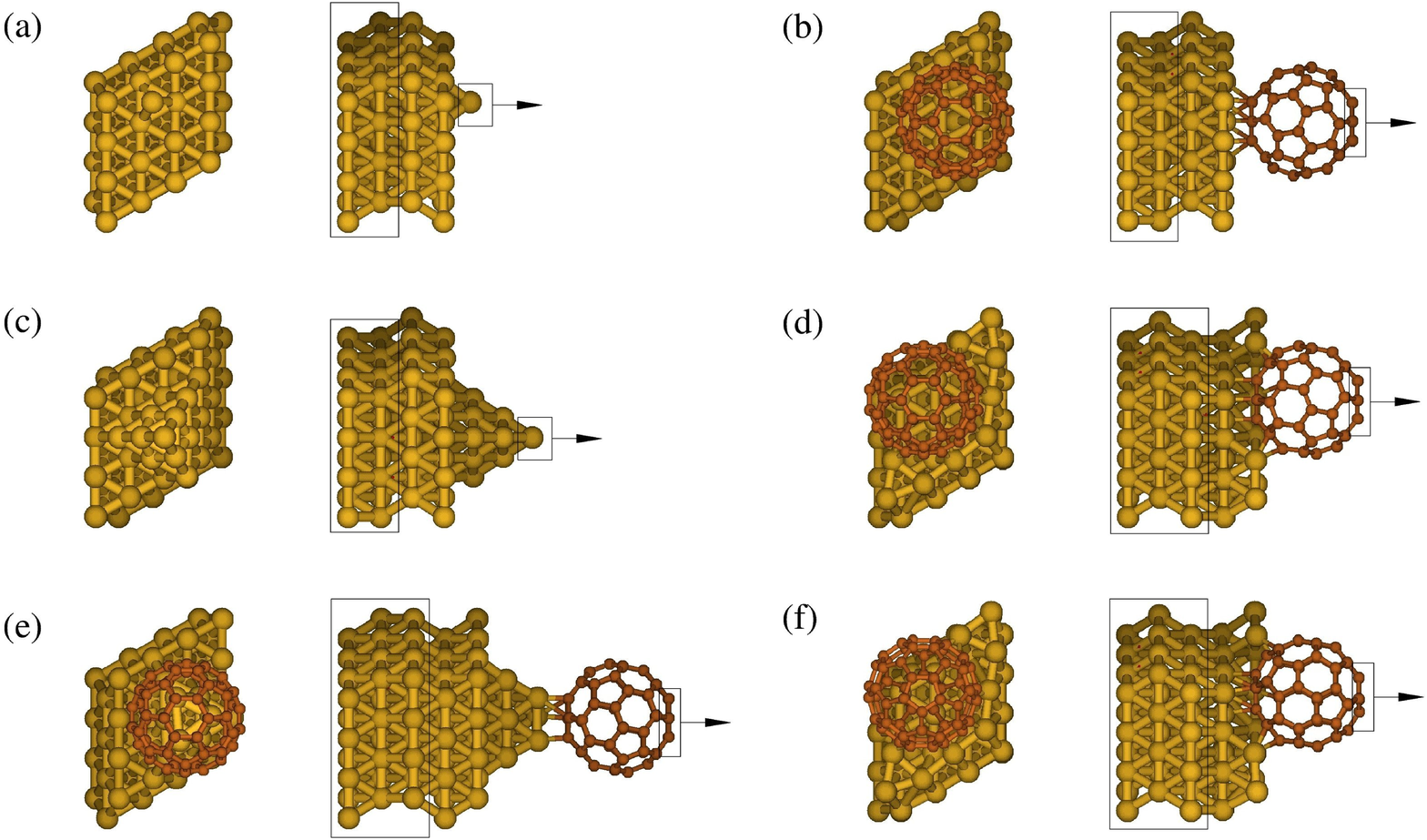}
\caption{Structures considered in the DFT calculations
(top and side views):
(a) Cu adatom on flat Cu(111), (b) \hex\ on flat Cu(111),
(c) pyramidal Cu tip, (d) \hex\ on reconstructed Cu(111),
(e) \hex-tip on Cu pyramid, (f) \pen\ on reconstructed Cu(111).
 The boxed atoms in the ($4\times4$)-Cu(111) slab are kept fixed at bulk
coordinates while the other degrees of freedom are relaxed. The elasticities
are estimated from the energy increase associated with displacing the top-most atoms (box with arrow)
perpendicular to the surface film while allowing the remaining degrees of freedom to relax.}
\label{fig:setup}
\end{figure}

In order to estimate the elasticities related to the experimental sample and tip sides we considered the generic structures shown in Fig.~\ref{fig:setup}.
Calculations were performed using the SIESTA \cite{Soler.02} pseudopotential density functional theory method with
the PBE-GGA exchange-correlation functional \cite{Perdew.96}, a 400 Ry mesh cutoff, and a $2\times2$ Monkhorst-Pack \cite{Monkhorst1976} $k$-mesh.
The Fermi surface was treated by a second-order Methfessel-Paxton \cite{Methfessel1989} scheme with an electronic temperature of 300 K.
The basis set consisted of default double-zeta plus polarization (DZP) orbitals for C and Cu atoms generated with an energy shift of 0.01 Ry.
The force tolerance for the structural optimizations was 0.02 eV/{\AA}. The total energies from SIESTA were not corrected for basis set superposition errors.
Two different lattice constants for the Cu crystal ($a = 3.62$ {\AA} and $a = 3.70$ {\AA}) were considered to confirm that the results do not depend sensitively on this parameter.

\appendix
\setcounter{section}{1}
\begin{figure}[h]
\centering
\includegraphics[width=0.7\textwidth,clip=]{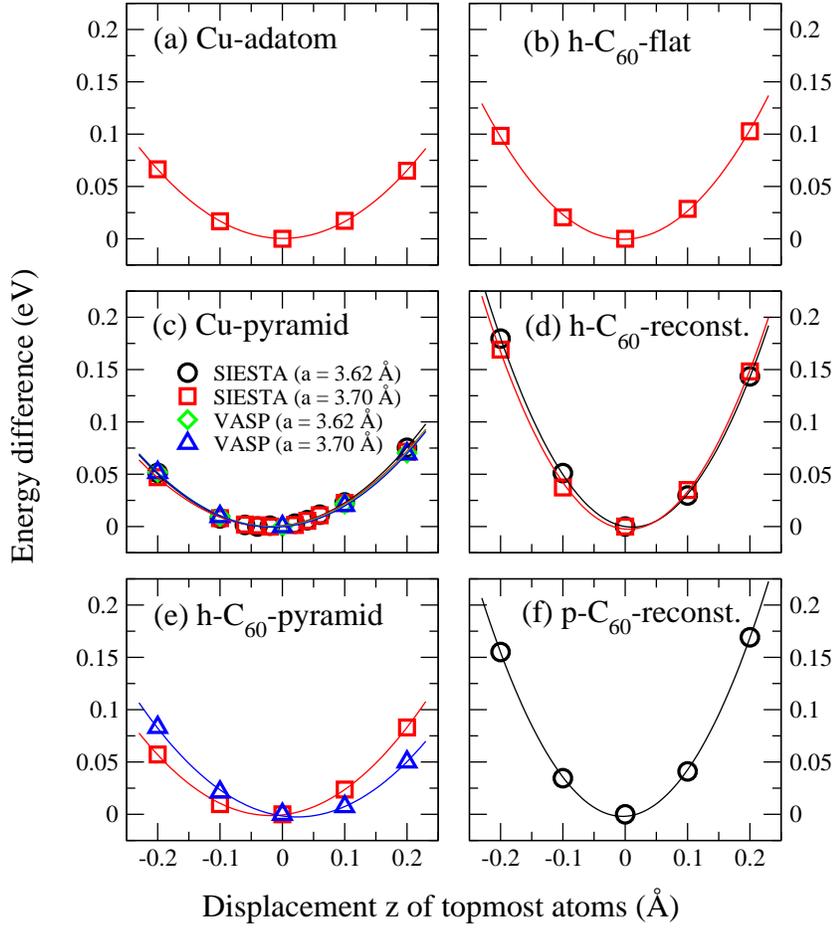}
\caption{Energy differences versus displacement $z$ of the topmost atoms
for the systems shown in Fig.~\ref{fig:setup}. The relaxed geometry corresponds
to $z = 0$ {\AA} whereas tensile strain is exerted on the system for $z > 0$ {\AA}. The solid lines are
quadratic fits to the calculated data points. The corresponding fitted spring constants are
reported in Tab.~\ref{tab:springs}.}
\label{fig:energetics}
\end{figure}
\begin{figure}[h]
\centering
\includegraphics[width=0.9\textwidth,clip=]{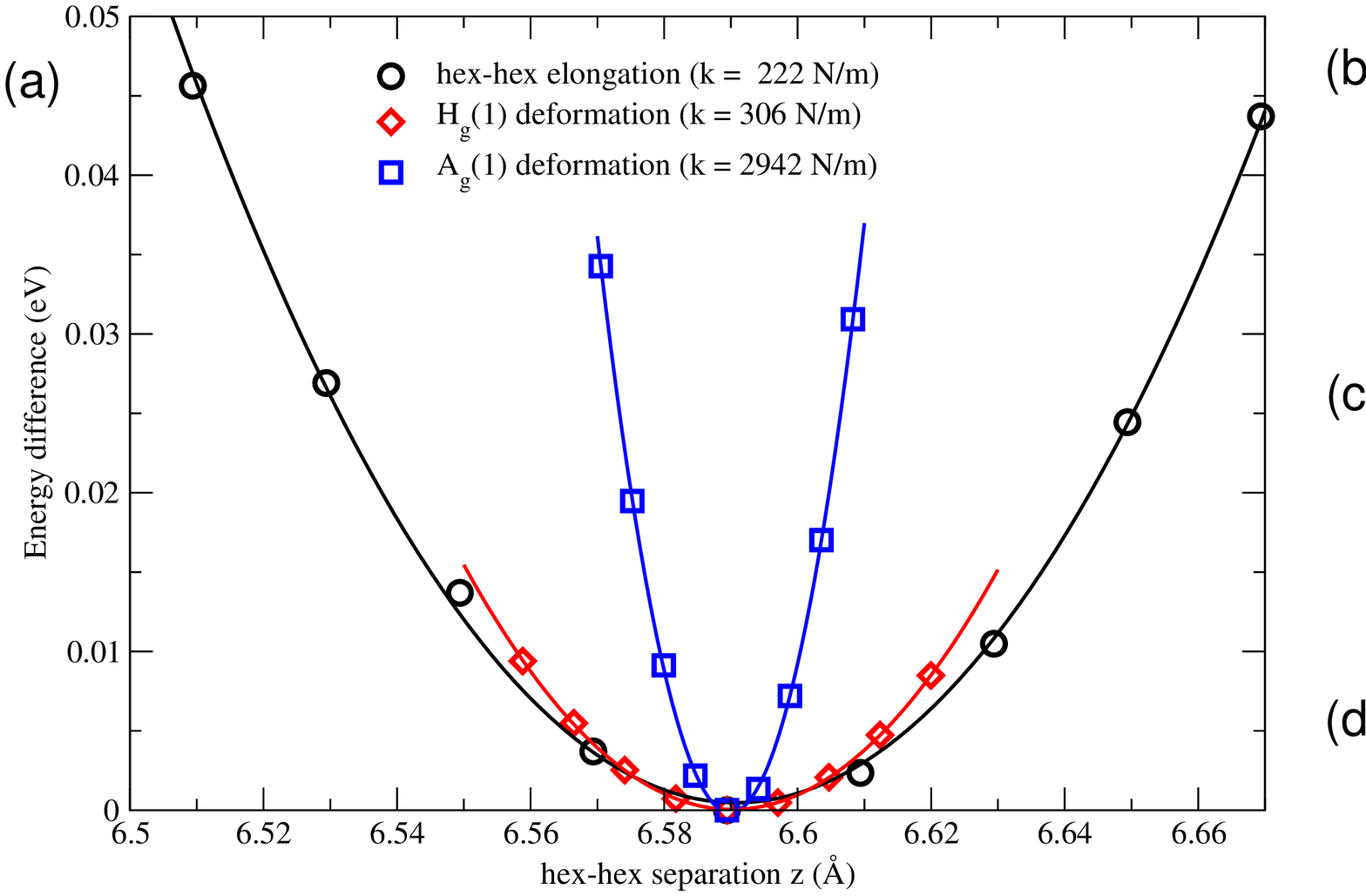}
\caption{(a) Energy differences versus hexagon-hexagon distance $z$ for different deformations of an isolated \C\ molecule.
The solid lines are quadratic fits to the calculated data points.
(b) Fixing the distance $z$ between opposite hexagons while relaxing all other degrees of freedom (black circles).
(c) H$_g$(1) vibrational mode of \C\ (compression/expansion, $\hbar\omega=31$ meV, red diamonds).
(d) A$_g$(1) vibrational mode of \C\ (isotropic deformation, $\hbar\omega=60$ meV, blue squares).}
\label{fig:isolatedC60}
\end{figure}

The computational procedure consisted of the following steps: (1) Relaxation of initial geometry with
the boxed atoms in the Cu slab (Fig.~\ref{fig:setup}) kept fixed at bulk coordinates while the other degrees of freedom fully relaxed.
(2) Displacement of the topmost atoms
(box with arrow in Fig.~\ref{fig:setup}) perpendicular to the surface film while relaxing all remaining degrees of freedom.

\appendix
\setcounter{section}{1}
\begin{table*}[h]
 \centering
\begin{tabular}{llccc}\hline\hline
Figure & System      & $a$  & $k_\textrm{SIESTA}$ & $k_\textrm{VASP}$ \\
&             & (\AA)  & (N/m)      & (N/m) \\
\hline
\ref{fig:energetics}(a) & Cu-adatom     &  3.70  &  52  (55)  & \\
\ref{fig:energetics}(b) & \hex-flat     &  3.70  &  81  (78)  & \\
\ref{fig:energetics}(c) & Cu-pyramid    &  3.62  &  50  (44)  &  49 (51) \\
                                          &               &  3.70  &  47  (45)  &  49 (47) \\
\ref{fig:energetics}(d) & \hex-reconst  &  3.62  &  129 (129) & \\
                                          &               &  3.70  &  129 (112) & \\
\ref{fig:energetics}(e) & \hex-tip      &  3.70  &  57  (53)  &  54 (43)\\
\ref{fig:energetics}(f) & \pen-reconst  &  3.62  &  131 (118) & \\
\ref{fig:isolatedC60}   & Isolated \C\   &        &  222 (240)\\
\hline\hline
\end{tabular}
\caption{Calculated spring constants $k_\textrm{SIESTA}$ and $k_\textrm{VASP}$ for the systems shown in Fig.~\ref{fig:setup} using the SIESTA and VASP codes, respectively.
The values in parentheses are derived from the second derivative of 5th order polynomial fits
evaluated at the energy minimum.
Two different lattice constants ($a = 3.62$ {\AA} and $a = 3.70$ {\AA}) were considered for the Cu crystal.
For comparison the stiffness of an isolated \C\ from Fig.~\ref{fig:isolatedC60} is included.}
\label{tab:springs}
\end{table*}
The energy costs associated with the deformations with respect to the displacement $z$ of the topmost atoms are shown in
Fig.~\ref{fig:energetics} with $z=0\,\text{\AA}$ being the relaxed junction. For $z > 0\,\text{\AA}$, tensile strain is exerted on the system.
Quadratic (as well as fifth order) fits to these energy differences yield the effective spring constants in Tab.~\ref{tab:springs}.
It should be noted that the elasticities in Tab.~\ref{tab:springs} represent in fact upper bounds because of the limited size of the unit cell.
In the real system more atoms respond to the pull on the topmost atoms, which leads to a smaller effective spring constant.

To test the accuracy of the SIESTA calculations, selected checks with the VASP code based on a plane-wave basis and the projector-augmented wave method (PAW) \cite{VASP1,VASP2,VASP3} were also performed.
We used PBE-GGA \cite{Perdew.96}, at least 400 eV energy cutoff, a $2\times2$ (or $3\times3$) Monkhorst-Pack \cite{Monkhorst1976} $k$-mesh, first-order Methfessel-Paxton \cite{Methfessel1989} scheme with 0.05 eV smearing width, and 0.02 eV/{\AA} force tolerance.
As seen in Fig.~\ref{fig:energetics} and Tab.~\ref{tab:springs}, the two codes yield similar estimates of the elasticities.

To determine effective elasticities for the combined elasticity of tip and sample the springs from Tab.~\ref{tab:springs} are added in series.
In this way, for a Cu-adatom [Fig.~\ref{fig:setup}(a)] or for a sharp pyramidal Cu-tip [Fig.~\ref{fig:setup}(c)] in contact with a \hex\ on
reconstructed Cu(111) [Fig.~\ref{fig:setup}(d)], we obtain $k_\mathrm{eff} = (1/k_\mathrm{tip}+1/k_\mathrm{surf})^{-1} \approx (32-39) \,\mathrm{N/m}$.
In the case of the contact between a Cu tip [Fig.~\ref{fig:setup}(a) or (c)] and a \hex\ on the flat Cu(111) [Fig.~\ref{fig:setup}(b)],
an effective elasticity of $k_\mathrm{eff} \approx (29-33) \,\mathrm{N/m}$ is obtained.
Due to the reduced number of bonds of \hex\ on flat Cu(111) in comparison with \hex\ on reconstructed Cu(111), $k_\mathrm{eff}$ is smaller.
Similarly, for the \C-tip [Fig.~\ref{fig:setup}(b) or (e)] in contact with a \hex\ on reconstructed Cu(111), we estimate $k_\mathrm{eff} \approx (31-50) \,\mathrm{N/m}$.

We also analyzed the stiffness of an isolated \C\ molecule with SIESTA, cf.~Fig.~\ref{fig:isolatedC60}. By controlling
the distance between two opposing hexagons while relaxing all other degrees of freedom we obtain an effective stiffness of
the molecule of $k_\textrm{eff} = 222$ N/m. A comparable (but slightly larger) stiffness of $k_\textrm{Hg1} = 306$ N/m is obtained when considering only a deformation
along the characteristic elongation/compression H$_g(1)$ vibrational mode [Fig.~\ref{fig:isolatedC60}(c)].
Deforming along the isotropic A$_g(1)$ vibrational mode [Fig.~\ref{fig:isolatedC60}(d)] yields a much larger
stiffness ($k_\textrm{Ag1} = 2942$ N/m).

\section{Elasticity model}

\begin{figure}[ht]
\centering
\includegraphics[width=150mm,clip=]{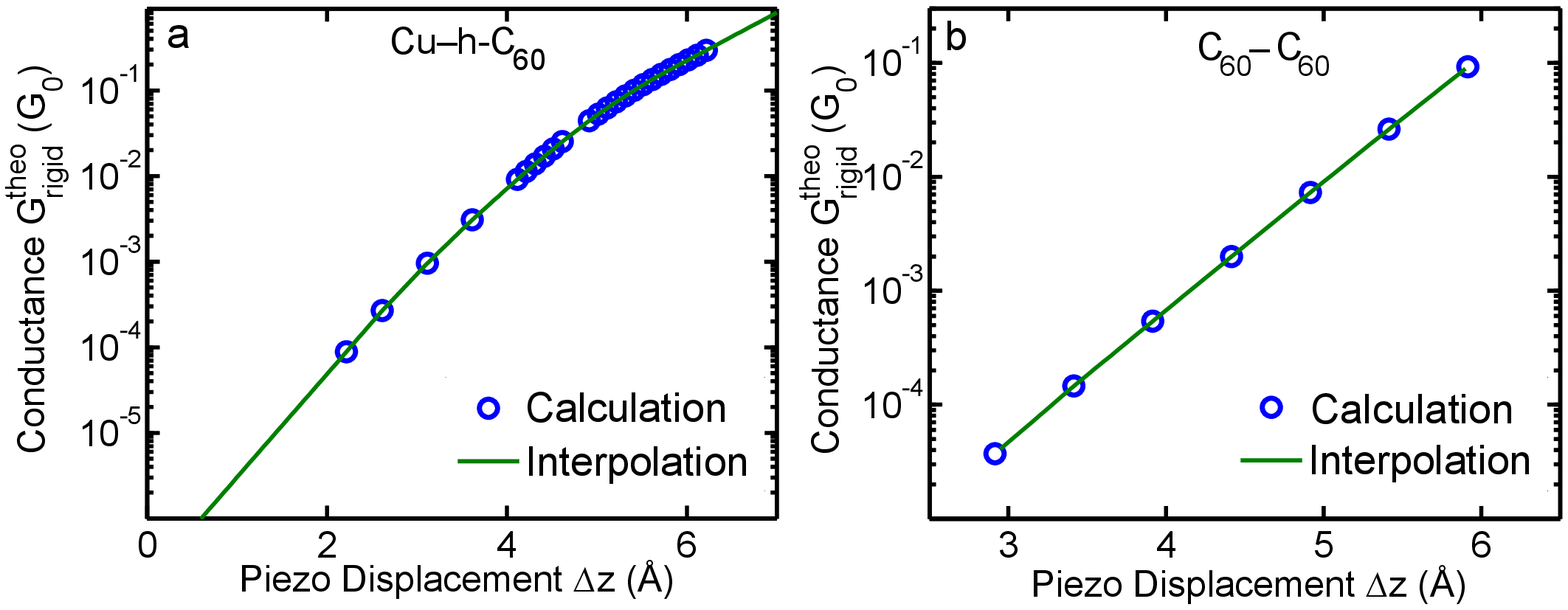}
\caption{Interpolated calculated conductance $G_\text{rigid}^\text{theo}$ as a function of the piezo displacement $\Delta z$ for the (a) Cu--\hex\ and (b) \C\--\C\ contact. For the Cu--\hex\ contact, $\Delta z =0\,\text{\AA}$ corresponds to a distance between the topmost tip atom and the topmost \C\ hexagon of $8\,\text{\AA}$. For the \C--\C\ contact, $\Delta z =0\,\text{\AA}$ corresponds to a \C--\C\ center distance of $14.9\,\text{\AA}$.}
\label{fig:Theo}
\end{figure}

The influence of the junction deformation on the conductance is described as follows.
First, the conductance of a rigid system as a function of the piezo displacement, $G_\text{rigid}^\text{theo} (\Delta z)$, is calculated for Cu--\hex\ and \C--\C\ contacts (Figs.~\ref{fig:Theo}(a) and (b)).
$\Delta z =0$ corresponds to a distance between the topmost tip atom and the topmost \C\--hexagon of $8\,\text{\AA}$ in the Cu--\hex\ contact and to a \C--\C\ center distance of $14.9\,\text{\AA}$ in the \C--\C\ contact.
Further calculational details can be found in Ref.\,\cite{Schull2009}.

Next, we assume that the deformation $d$ depends linearly on the experimentally determined force, i.~e. $d = F(\Delta z)/k$ with $k$ describing the effective stiffness of the junction.
The theoretical conductance of the deformed junction, $G_\text{deform}^\text{theo} (\Delta z)$, depending on $k$ is given by:
\begin{equation}
G_\text{deform}^\text{theo} (\Delta z) =G_\text{rigid}^\text{theo} \Big(\Delta z-\frac{F(\Delta z)}{k}\Big)\,\,.\label{seq1}\end{equation}
The interpolated calculated conductance $G_\text{deform}^\text{theo}$ is then fitted to the experimental conductances $G(\Delta z)$.
As the absolute tip--sample distance was unknown in the experiments, an arbitrary shift on the abscissa for $G_\text{rigid}^\text{theo} (\Delta z)$ is allowed.
This shift is already comprised in Fig.~\ref{fig:Theo}, so that the $\Delta z$ axes in Fig.~\ref{fig:Theo}(a) and (b) correspond to the axis in Fig.~3.

\appendix
\setcounter{section}{2}

Fits for the Cu--\hex\ and \C--\C\ contacts shown in Fig.~3(a) and to similar data calculated for other microscopic tips yield $k \approx \left( 16-37 \right)\,\text{Nm}^{-1}$ and $k\approx \left(13-24\right)\,\text{Nm}^{-1}$, respectively.
Compared to the values from the DFT calculations in appendix A, the effective spring constants qualitatively agree, but are smaller by a factor of 2.
This deviation is not unexpected since the calculated elasticities are  upper bounds.
Furthermore, the model neglects the formation of chemical bonds between tip and sample, which are expected to weaken the effective spring constant.

\end{document}

\title{Force and conductance during contact formation to a \C\ molecule\\ - Supplementary Data -}

\author{Nadine Hauptmann$^1$, Fabian Mohn$^2$, Leo Gross$^2$, Gerhard Meyer$^2$, Thomas Frederiksen$^3$, and Richard Berndt$^1$}
\address{$^1$Institut f\"{u}r Experimentelle und Angewandte Physik, Christian-Albrechts-Universit\"{a}t zu Kiel, D-24098 Kiel, Germany\\}
\address{$^2$IBM Research -- Zurich, CH-8803 R\"{u}schlikon, Switzerland\\}
\address{$^3$Donostia International Physics Center (DIPC), E-20018 Donostia-San Sebasti$\acute{a}$n, Spain}
\ead{hauptmann@physik.uni-kiel.de}

\maketitle

\section{Experimental details}

Frequency shift--distance curves were recorded as follows.
The current feedback was disabled and the tip was retracted from the surface by several angstroms.
The frequency shift $\Delta f$ and the time-averaged tunneling current $I_\text{avg}$ were recorded while approaching the tip towards the surface and during the subsequent retraction to its starting position.
Curves were acquired with 2000 data points in 60 to 120\,s.
Since no significant differences were observed between the approach and retraction data, modifications of the tip apex or the molecular orientation during measurements can be excluded.
In addition, STM images taken before and after current--distance measurements showed no changes.
Furthermore, the intrinsic energy dissipation of the tuning fork did not change significantly during approach and retraction of the tip.
\begin{figure}[ht]
\centering
\includegraphics[width=150mm,clip=]{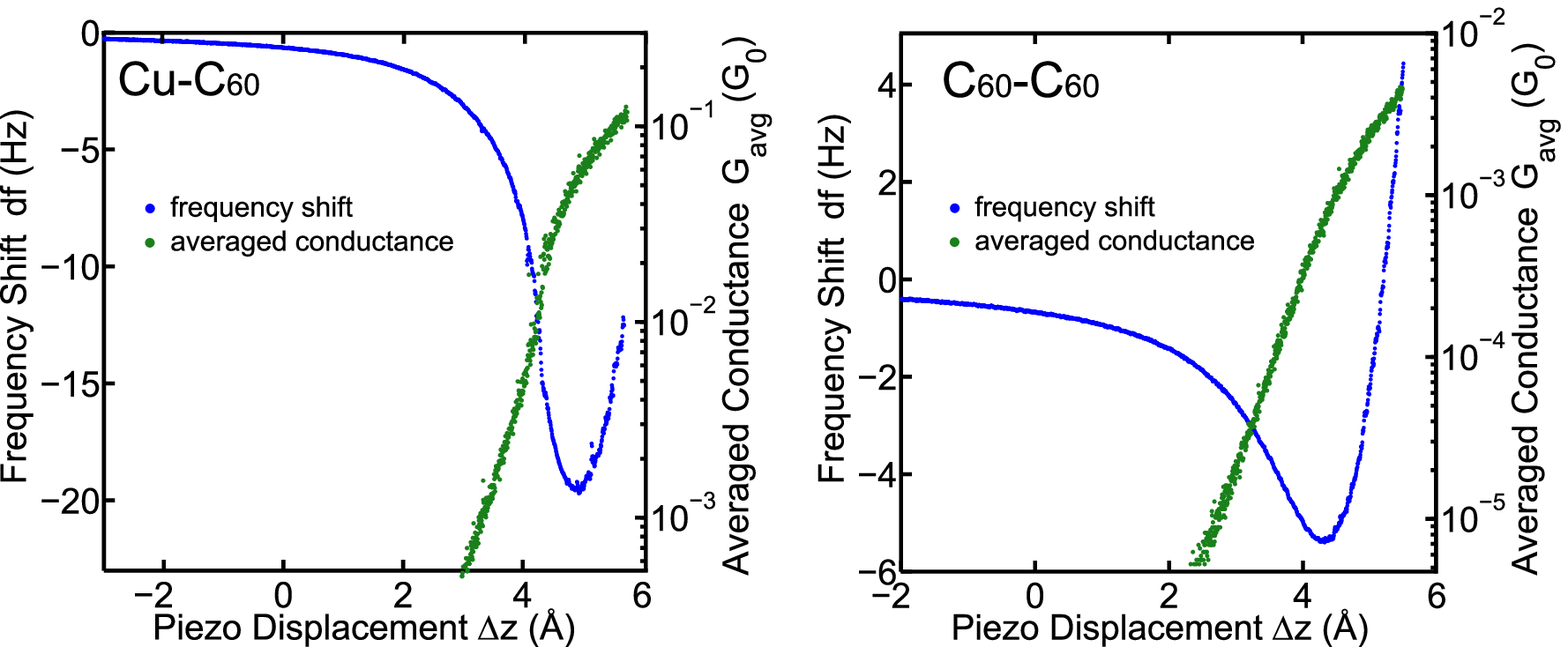}
\caption{Frequency shift $df$ and time-averaged conductance $G_{avg}$ versus the piezo displacement $\Delta z$ for the  Cu--\C\ and \C--\C\ contacts shown in Fig.~1(c) and Fig.~2(h), respectively. The oscillation amplitudes are $A=(3\pm0.2)\,\text{\AA}$ for the Cu--\C\ contact and $A=(2.5\pm0.2)\,\text{\AA}$ for the \C\--\C\ contact.}
\label{fig:RawData}
\end{figure}
As the cantilever, one prong of a tuning fork was used (spring constant $k\approx 1800\,\text{N/m}$) while the second prong was fixed on the piezo scanner.
Measurements were performed at oscillation amplitudes between $A=1\,\text{\AA}$ and $A=3\,\text{\AA}$.
The amplitude $A$ corresponds to half of the peak-to-peak distance of the oscillation.
The uncertainty of the calibration of the used amplitude is $\approx 7\text{\%}$. This margin is essentially due to the uncertainty of the calibration of the z-piezo, where $\approx 5\text{\%}$ are routinely achieved for small displacements.
The bandwidth of the transimpedance amplifier for the tunneling current varies between $7$ and $200\,\text{kHz}$ depending on the used amplifier gain.
The current data was further filtered by a numerical low-pass filter with a cut-off frequency of $8\,\text{kHz}$.

\subsection{Extraction of the short-range force}
\begin{figure}[ht]
\centering
\includegraphics[width=100mm,clip=]{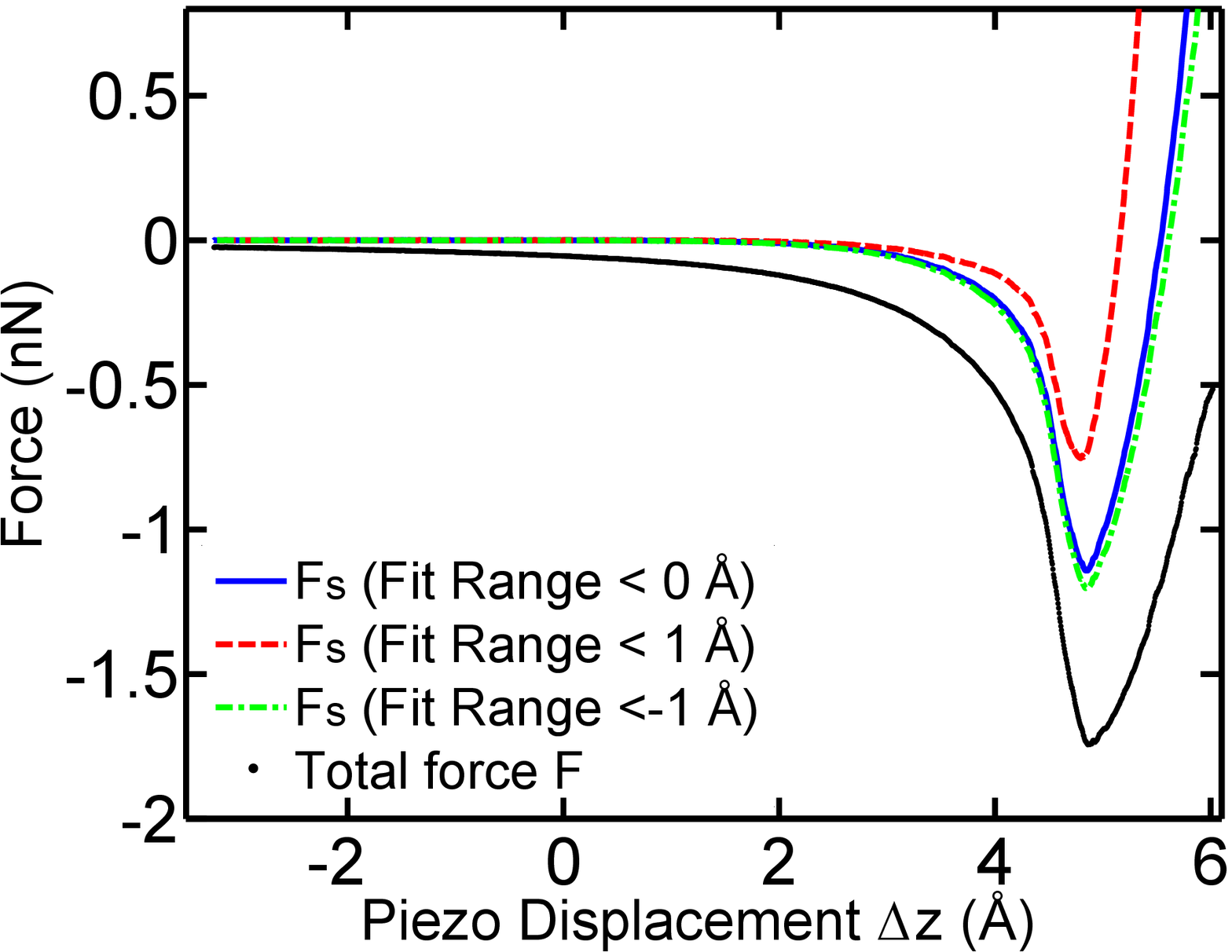}
\caption{Total force $F$ (dots) and short-range forces $F_s$ (solid, dashed, and dashed-dotted lines) versus the piezo displacement $\Delta z$. The three $F_s$ curves are extracted by three long-range force fits with different fit ranges.}
\label{fig:longRange}
\end{figure}

The short-range force $F_s$ can be obtained by subtracting the long-range van-der-Waals force $F_l$ from the total force $F$. We use a power law to approximate $F_l$ by $F_l(\Delta z) = a (\Delta z_0 - \Delta z)^b$.
The fit parameters depend on the range of the fit.
E.g., for the fit range $\Delta z \leq 0\,\text{\AA}$ we find $a=-5.5\,\text{nN}/\text{\AA}^{b}$, $b=-2.3$, and $\Delta z_0=7.5\,\text{\AA}$.
For $\Delta z \leq -1\,\text{\AA}$ and $\Delta z \leq 1\,\text{\AA}$ we obtain $a=-6.5\,\text{nN}/\text{\AA}^{b}$, $b=-2.34$, $\Delta z_0=7.8\,\text{\AA}$ and $a=-2\,\text{nN}/\text{\AA}^{b}$, $b=-1.9$, $\Delta z_0=6.3\,\text{\AA}$, respectively.
Figure~\ref{fig:longRange} shows the influence on $F_s$ for the different fit ranges.
The absolute value of $F_s$ can change by a factor of 2 depending on the fit range.
However, the displacement $\Delta z$ where contact occurs is only weakly influenced by the different long-range force fits.
It shifts by $\approx 0.12\,\text{\AA}$ at most.
Therefore we did not use $F_s$ for a quantitative comparison.
Rather we compare the more robust point of maximal attraction with the point of contact.

\section{Simulation of images recorded with \C-tips}

\begin{figure}[ht]
\centering
\includegraphics[width=130mm,clip=]{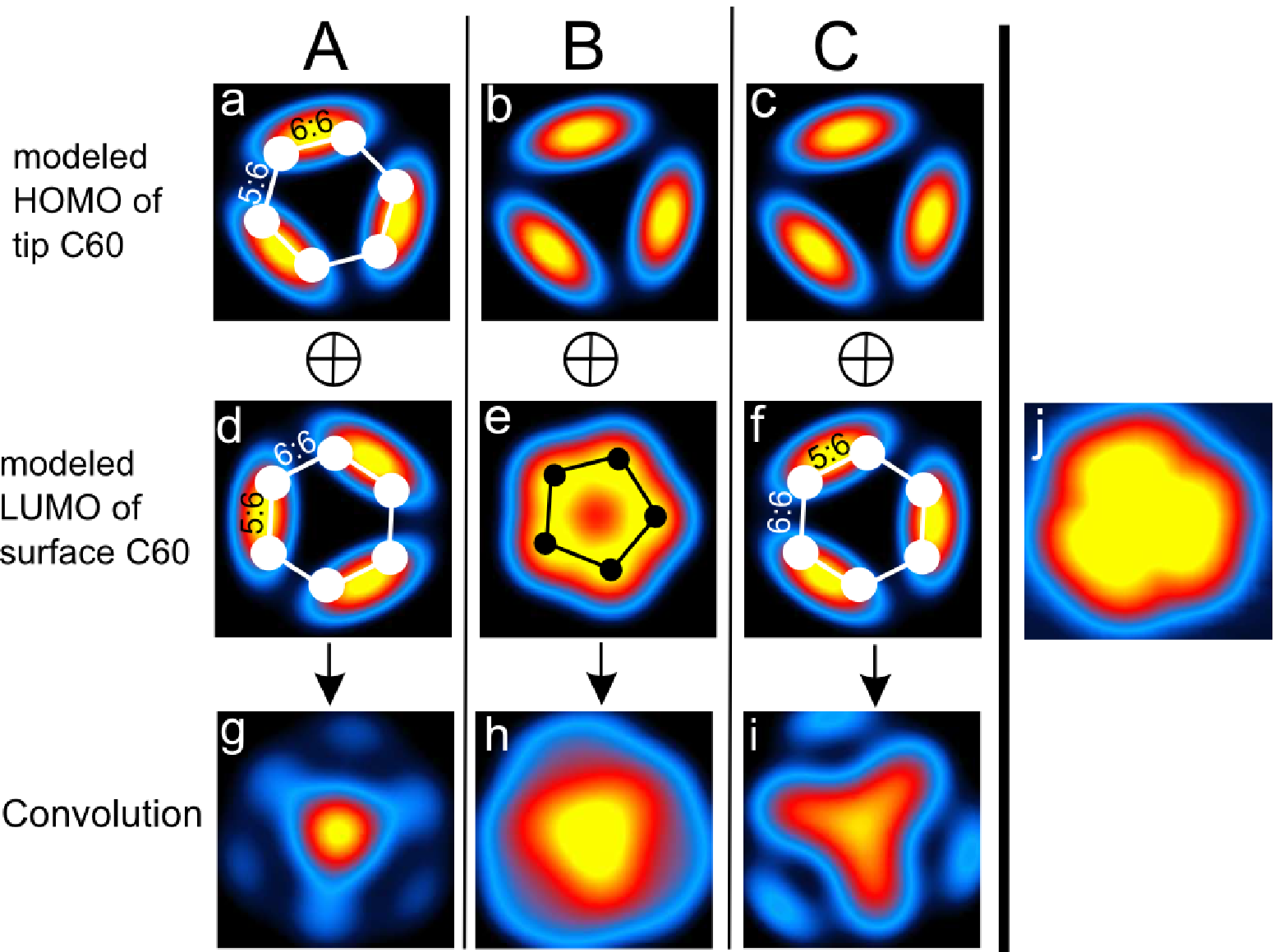}
\caption{(a--c) Model LDOS of the HOMO of a \hex-tip hexagon as viewed from the tip.
(d--f) Model LDOS of the LUMO of \hex\ and \pen\ on the surface. (g--i) Two-dimensional convolution of tip and sample LDOS\@. White hexagons and a black pentagon indicate the molecular orientation as viewed from the tip. A, B, and C refer to the surface regions defined in Figs.~1(b) and Fig.~2(b). (j) Constant-current STM image ($-2\,\text{V}$, $0.26\,\text{nA}$, $1.7\times 1.6\,\text{nm}^2$) of a Cu cluster with a \C\ tip.}
\label{fig:conv}
\end{figure}

The highest occupied molecular orbitals (HOMO) of \C\ exhibit a large local density of states (LDOS) at the 6:6 bonds, which is known from DFT calculations.
Motivated by this result, we modeled the HOMO of a \hex-tip by a threefold-symmetric pattern of two-dimensional Gaussians [Fig.~\ref{fig:conv}(a)-(c)].
For simplicity it was assumed that only the LDOS of the hexagon which is closest to the surface contributes to the tunneling current.
The same approach was used to mimic the lowest unoccupied molecular orbital (LUMO) of \hex\ and \pen\ on the substrate, which are most intense at the 5:6 bonds [Fig.~\ref{fig:conv}(d--f)].
As images of adsorbed \hex\ recorded with a Cu-tip reveal the positions of the bonds, the relative orientation of \C\ at the tip and the surface were determined from the experiment. For \pen\ the situation is less favorable, because the molecule appears circularly symmetric when using a Cu tip.
A two-dimensional convolution of these patterns [Figs.~2(e--g) and Figs.~\ref{fig:conv}(g--i)] reproduces the experimental images of \hex\ qualitatively very well.
The orientation of the \C\ molecule attached to the tip was determined by scanning a Cu cluster, consisting of a few Cu atoms, with the \C\ tip. Fig.~\ref{fig:conv}(j) shows the experimental data ($V=-2\,\text{V}$).
This image is rotated by 180\degree (about an axis perpendicular to the image plane) compared to that of a \C\ on the sample surface scanned with a metal tip.